\documentclass[10pt]{article}

\usepackage{amsmath}
\usepackage{amssymb}
\usepackage{graphicx}
\usepackage{cite}
\usepackage{color}
\usepackage{rotating}
\usepackage{amsmath}
\usepackage{amssymb}
\usepackage{amsfonts}
\usepackage{color}
\usepackage{float}
\usepackage{caption}
\usepackage{subcaption}

\definecolor{light-gray}{gray}{0.90}

\begin{document}

\begin{flushleft}

{\LARGE \textbf{\textit{In silico} tumor control induced via alternating immunostimulating and immunosuppressive phases}}
\\
\vspace{5mm}

A. I. Reppas$^{1}$, J. C. L. Alfonso$^{1}$ and H. Hatzikirou$^{1, \dagger}$
\\
\vspace{3mm}

$^{1}$ Center for Advancing Electronics, Technische Universit\"{a}t Dresden, 01062 Dresden, Germany.
\\
\vspace{3mm}

$^{\dagger}$ Corresponding author: haralambos.hatzikirou@tu-dresden.de

\end{flushleft}


\section*{Abstract}

Despite recent advances in the field of Oncoimmunology, the success potential of immunomodulatory therapies against cancer remains to be elucidated. One of the reasons is the lack of understanding on the complex interplay between tumor growth dynamics and the associated immune system responses. Towards this goal, we consider a mathematical model of vascularized tumor growth and the corresponding effector cell recruitment dynamics. Bifurcation analysis allows for the exploration of model's dynamic behavior and the determination of these parameter regimes that  result in immune-mediated tumor control. Here, we focus on a particular tumor evasion regime that involves tumor and effector cell concentration oscillations of slowly increasing and decreasing amplitude, respectively. Considering a temporal multiscale analysis, we derive an analytically tractable mapping of  model solutions onto a weakly negatively damped harmonic oscillator. Based on our analysis, we propose a theory-driven intervention strategy involving immunostimulating and immunosuppressive phases to induce long-term tumor control.


\section*{Introduction}

The immune system is widely recognized for its capacity to detect and destroy cancer cells, as well as to prevent tumor recurrence maintaining an immunological memory \cite{Finn2012}. Indeed, every known innate and adaptive immune effector mechanism has been reported that participates in tumor recognition and rejection \cite{Dranoff2004}. However, experimental observations support that tumorigenic processes themselves can promote immunosuppression or immune tolerant states that facilitates neoplastic growth and progression \cite{Stewart2008, Zou2005}. Cancer cells employ diverse strategies to inhibit or block immune responses, including tumor-induced impairment of antigen presentation, secretion of immunosuppressive cytokines or expression of surface molecules, as well as production of diverse pro-apopoptic factors \cite{Rabinovich2007}. Nevertheless, there are clinical and preclinical evidences supporting that activation of the innate antitumor immunity can result in tumor regression and provide therapeutic benefits \cite{Ghirelli2013}.

The main goal of oncoimmunology is to strengthen the immune system's innate ability to combat and kill cancer cells by enhancing the effectiveness of the immune responses. Among the different immunotherapeutic techniques are checkpoint inhibitors, immune response modifiers (cytokines), monoclonal antibodies and vaccines \cite{Cheever2008}. Passive and active immunotherapy has been successfully applied to the treatment of a wide variety of human cancers \cite{Rosenberg2014} and holds promise of a lifelong cure \cite{Finn2008}. However, tumor-induced immunosuppression still represents a major obstacle to effective cell-mediated immunity and immunotherapy \cite{Rabinovich2007, Stewart2008}. Accordingly, more insights into the main mechanisms associated with immune responses based on tumor specific features are required to obtain successful therapeutic outcomes with immunomodulatory strategies. Despite years of research devoted to understand the underlying mechanisms of immune-tumor interactions, there are still many unanswered questions. In particular, those related with the impact of tumor-associated vascularization on immune responses, as well as determination of optimal and effective therapeutic protocols in cancer immunotherapy, are far from being completely elucidated.

Mathematical oncology is a valuable descriptive and predictive analytic framework to address such open questions. Continuum Mechanics concepts have been widely considered to investigate tumor growth and therapy implications \cite{ambrosi2002mechanics, byrne2003modelling, jain2014role, ramirez2015action, ramirez2015mathematical}. In addition, several mathematical models of tumor growth, where some forms of the immune dynamics are often included, have been also extensively studied in the last years \cite{Arciero2004, Bellomo1999, Bellomo2000, Galach2003, Matzavinos2004, DOnofrio2005, Mallet2006, AlTameemi2012, RobertsonTessi2012, Rihan2014}. Clinical data evidence that cancer cells can survive in a undetectable dormant state for extended periods of time \cite{Koebel2007}, which has been also predicted by several models of tumor-immune cell interactions \cite{Matzavinos2004, DOnofrio2005, DOnofrio2007, Caravagna2010, AlTameemi2012}. However, the neoplasm develops diverse strategies to circumvent the anti-tumor action of the immune system \cite{Pardoll2003, Dunn2004}. In particular, this equilibrium state can be disrupted by different events affecting the immune system which could result in tumor regrowth \cite{Dunn2004}. Sustained oscillations by the immune system have been observed both in its healthy state and pathological situations \cite{Stark2007, Coventry2009, Caravagna2010}. Therefore, the presence of an immune component in mathematical modeling has been described crucial for reproducing clinically observed phenomena such as tumor dormancy, oscillations in tumor size and spontaneous tumor regression \cite{Kuznetsov1994, Kirschner1998, DOnofrio2006, Pillis2006, Caravagna2010, DOnofrio2010}. Among the several reviews on the subject are those covering mathematical models of tumor growth mainly focused on the cancer and immune system interactions \cite{Araujo2004, Nagy2005, Byrne2006, Roose2007, Martins2007, Bellomo2008, Chaplain2008, Eftimie2011, Wilkie 2013}. The mathematical modeling of the entire immuno-oncology dynamics is an enormously difficult and complex task. In consequence, models describing interactions between growing tumors and immune dynamics should focus on the crucial factors that are known to allow tumor escape from immuno-surveillance. Tumor-induced angiogenesis is a crucial mechanism for cancer survival and proliferation, allowing a continuous supply of oxygen and nutrients needed for tumor growth and progression \cite{Ruoslahti2002, Ferrara2005, Farnsworth2014}. However, the effectiveness of antitumor immune responses is associated with the functional levels of the tumor blood vessels, which allow a wider range of effector cell types to penetrate the tumor bulk and further exterminate cancer cells \cite{Fridman2012, Junttila2013}. These opposing effects demand for a mathematical model of vascularized tumor growth that allows to explore the therapeutic potential of immunomodulatory interventions when innate immune responses are insufficient for long-term tumor control.

The work herein reported intends to yield new insights in the potential of immunomodulatory interventions for cancer therapy. To this end, we propose a tumor-effector cell model based on well-known biological assumptions that combines a model of radially symmetric tumor growth with an immune cell recruitment model \cite{Greenspan1976, Cristini2003, Kuznetsov1994, DOnofrio2005}. The main feature of our model is the modeling of the interplay between functional tumor-associated vasculature and effector cell dynamics as described in \cite{Hatzikirou2015}. Model results predict that, depending on the functional tumor vascularization degree and effector cell recruitment rate, long-term tumor control cannot be always reached. We particularly focus in such situations where tumors escape immuno-surveillance to suggest an optimized theory-driven therapeutic strategy against tumor growth. A temporal multiscale approach is then implemented to describe the tumor-immune system interactions, where an analytically tractable approximation of the cancer-effector cell dynamics is derived. We find that an efficient modulation of the immunostimulating and immunosuppressive phases could induce long-term tumor control.


\section*{Mathematical model description}
\label{sec: mathematicalmodel}

The present model describes the interactions between growing tumors and induced immune system dynamics. More precisely, the proposed tumor-effector cell model is a combination of a radial tumor growth model and an effector cell recruitment model originally proposed in \cite{Greenspan1976} and \cite{Kuznetsov1994} respectively, see also \cite{Cristini2003, Hatzikirou2012, DePillis2005, DOnofrio2005}. The main feature is the low dimensional modeling of the complex interplay between tumor-associated functional vasculature and immune recruitment dynamics as described in \cite{Hatzikirou2015}. This model can be interpreted as the temporal evolution of the average tumor radius, since radially symmetric growth is not a realistic behaviour. The system variables are the average tumor radius $R(t)$ and effector cell concentration $E(t)$ in the tumor vicinity. 


\subsection*{Radial tumor growth, $R(t)$}
\label{sec: radialtumorgrowth}

The temporal evolution of the average tumor radius is considered, where for simplicity invasive and diffusive tumor properties are not taken into account. The tumor is modeled as an incompressible fluid flowing through a porous medium, where tissue elasticity is simplified. The tumor-host interface is assumed to be sharp and cell-to-cell adhesive forces are modeled as a surface tension at that interface. The tumor expands as a mass whose growth is governed by a balance between cell birth (mitosis) and death (apoptosis and necrosis). The mitotic rate within the tumor is assumed to be linearly dependent on the nutrient concentration (oxygen, glucose, etc.) and is characterized by its maximal value $\lambda_M$ at the tumor-host interface. The death rate $\lambda_A$ is uniform within the tumor and constant in time. Moreover, we assume that the death rate $\lambda_A$ reflects the lump effect of apoptotic/necrotic processes and any other cell death inducing factor. The concentration of nutrients (e.g. oxygen or glucose) obeys a reaction-diffusion equation in the tumor volume, where nutrients are supplied from the tumor-associated functional vasculature and consumed by the tumor cells at a uniform consumption rate. 

To gain insight on the impact of vascularization in tumor growth and immune responses, we assume that the non-negative and dimensionless parameter $B$, where $0 \leq B \leq 1$, represents the net effect of functional tumor-associated vasculature on the tumor radius evolution. In the limit of avascular tumor growth $B = 0$, such tumor-effector interactions take place only at the tumor surface. At the other extreme, for $B = 1$ effector cells can potentially interact with any cancer cell within the tumor bulk. Moreover, we consider an intrinsic length scale $L_D$ representing the average length of nutrient gradient, i.e. supply, diffusion and consumption. 

The efficacy of immune killing depends on the ability of effector cells to penetrate the tumor bulk via the functional tumor-associated blood vessels. With improved vascularization, the effectors kill tumor cells not only on the surface of the tumor, but also further inside \cite{Huang2013}. We consider this process through a phenomenological scaling function $f(R,B) = \frac{R^{B - 1}}{R^{B - 1} + 1}\in[0,1]$, for $B \in [0,1)$, that models the penetration of effector cells in the tumor parenchyma through the existing functional vasculature. This function modulates the term related to tumor-effector cell interactions, such as killing of tumor cells due to effectors represented by a rate $c$. Such scaling functions have been also considered in the classical von Bertalanffy approach, and more recently in allometric models \cite{Herman2011}.

Taking these factors into account, we deduce the tumor radius $R(t)$ dynamic under the assumption of radial symmetry according to the following ODE equation:

\begin{equation}
\newcommand\hh{\vphantom{\lambda_M (1 - B) L_D \left( \frac{1}{\tanh(R / L_{D})} - \frac{L_{D}}{R} \right)}}
\label{eq:TE_1} \frac{dR}{dt} = \underbrace{\frac{1}{3} (\lambda_M B - \lambda_A) R \hh}_{\begin{subarray}{c} \vspace{1mm}\\ \mbox{net vascular} \vspace{1mm}\\ \mbox{tumor growth}\end{subarray}} + \underbrace{\lambda_M (1 - B) L_D \left( \frac{1}{\tanh(R / L_{D})} - \frac{L_{D}}{R} \right)}_{\begin{subarray}{c} \vspace{1mm}\\ \mbox{avascular} \vspace{1mm}\\ \mbox{tumor growth}\end{subarray}} -  \underbrace{c E R f(R,B) \hh}_{\begin{subarray}{c} \vspace{1mm}\\ \mbox{death due to} \vspace{1mm}\\ \mbox{effector cells}\end{subarray}}
\end{equation}

where the time coordinate $t$ has been omitted for notational simplicity, and $\lambda_M$, $\lambda_A$, $L_D$ and $c$ are non-negative constants.


\subsection*{Effector cell concentration, $E(t)$}
\label{sec: effectorcells}

We assume that effectors are recruited at a rate $r$ depending on tumor cells following the Michaelis-Menten kinetics, where $K$ is a positive constant denoting the concentration at which the immune recruitment is half-maximal. Effector cells die at a rate $d_0$, and become inactivated a rate $d_1$ due to their antitumor activity. In particular, the inactivation of effectors by tumor cells is modeled through the function $f(R,B)$ described above. As functional vascularization increases, effectors can kill cancer cells throughout the tumor bulk and tumor-immune cell interactions increase resulting in more inactivated immune cells. Moreover, innate immunity or base immuno-surveillance is represented as a minimum presence of active effector cells at any time given by an innate immune growth rate $\sigma$, even in the absence of tumor cells.

The resulting ODE equation for the effector cell concentration $E$ dynamic is given by:

\begin{equation}
\newcommand\ff{\vphantom{r \frac{R^3}{K + R^3} E}}
\label{eq:TE_2} \frac{dE}{dt} =  \underbrace{r \frac{R^3}{K + R^3} E}_{\begin{subarray}{c} \vspace{1mm}\\ \mbox{immune} \vspace{1mm}\\ \mbox{recruitment rate}\end{subarray}} - \underbrace{d_1 E R^{3} f(R,B) \ff}_{\begin{subarray}{c} \vspace{1mm}\\ \mbox{effectors} \vspace{1mm}\\ \mbox{inactivation rate}\end{subarray}} - \underbrace{d_0 E \ff}_{\begin{subarray}{c} \vspace{1mm}\\ \mbox{effectors} \vspace{1mm}\\ \mbox{death rate}\end{subarray}} + \underbrace{\sigma \ff}_{\begin{subarray}{c} \vspace{1mm}\\ \mbox{innate} \vspace{1mm}\\ \mbox{immunity rate}\end{subarray}}
\end{equation}

where the time coordinate $t$ has been omitted for sake of simplicity in the notation, and $r$, $K$, $d_0$, $d_1$ and $\sigma$ are non-negative constants.


\subsection*{Model parameterization}
\label{sec: parameterization}

Under the small tumor radius assumption, the very early tumor growth is always of exponential nature and does not depend on the vascularization effects, i.e. parameter $B$, \cite{Bru2003, Cristini2003, Bru2004, Hatzikirou2012}. Accordingly, we assume that this initial growth depends exclusively on the net proliferation rate $\lambda_p = (\lambda_M - \lambda_A)$ in the absence of adaptive antitumor immune responses at those stages of growth, see the first term of Eq.~(\ref{eq:TE_1}). Thus, considering experimental estimates of the growth rate at early phases of spheroids tumor growth for the mouse colon carcinoma cell line CT26 as $\lambda_p \approx 1.20$ day$^{-1}$ \cite{Alessandri2013, Hatzikirou2015} and the physiological plausible value $\lambda_M = 1/18 \,h = 1.34$day$^{-1}$ for CT26 murine cells \cite{Alessandri2013, Delarue2013, Hatzikirou2015}, we have that $\lambda_A \approx 0.14$day$^{-1}$. The characteristic nutrient diffusion length has been experimentally estimated that ranges between 0.2 and 0.3 mm \cite{Acker1984, Frieboes2006, Cristini2008, Hatzikirou2015}. The effector cell characteristic concentration is at the order of magnitude $10^5$ cells. The latter estimate is justified since the characteristic length scale of the system is at the order of 1.0 mm, and given that cells are commonly assumed with a diameter between 10$\mu$m and 20$\mu$m \cite{Delarue2013}, then for a volume of $1mm^3$ the concentration is at the order of magnitude $10^5$ cells. Moreover, we consider $c = 0.03$ cells$^{-1}$ days$^{-1}$ as measured from murine CT26 tumor growth experiments in \cite{Hatzikirou2015}, see also \cite{Kuznetsov1994, DOnofrio2005, DePillis2005}. The remaining parameter values $d_0 = 0.37$ day$^{-1}$, $d_1 = 0.01$ mm$^{-3}$ days$^{-1}$, $K = 2.72$ mm$^3$ and $\sigma = 0.13 \times 10^5$ cells days$^{-1}$ are considered from previously reported experimental data \cite{Kuznetsov1994, DOnofrio2005, DePillis2005, Su2009, DOnofrio2012}, and properly rescaled to the magnitudes and units considered in our model. Through a parameter sensitivity analysis, we explore the effects on tumor growth of the effector cell recruitment rate $r$ and functional vascularization degree $B$ for different choices of the initial tumor radius $R_0$ and concentration of effector cells $E_0$.

For convenience of the reader, we summarize in Tab.~\ref{tab:params} the parameter values used in numerical simulations of the tumor-immune dynamics considering the effect of tumor-associated vasculature.

\begin{table}[H]
\centering
\begin{tabular}{lllll}
\hline
Description & Parameter & Value & Units & Sources \\
\hline
 Tumor mitotic rate                                       & $\lambda_M$ & 1.34 & days$^{-1}$ & \cite{Alessandri2013, Delarue2013, Hatzikirou2015} \\
 Tumor death rate                                        & $\lambda_A$ & 0.14 & days$^{-1}$ & \cite{Alessandri2013, Delarue2013, Hatzikirou2015} \\
 Characteristic nutrient diffusion length       & $L_D$ & [0.2 - 0.3] & mm & \cite{Acker1984, Frieboes2006, Cristini2008, Hatzikirou2015} \\
 Tumor cell kill by effectors                          & $c$ & 0.03 & cells$^{-1}$ days$^{-1}$ & \cite{Kuznetsov1994, DOnofrio2005, DePillis2005, Hatzikirou2015} \\
 Tumor volumen where $r$ is half-maximal & $K$ & 2.72 & mm$^3$ & \cite{Kuznetsov1994, DOnofrio2005, DePillis2005, DOnofrio2012} \\
 Effectors inactivation rate by tumor cells    & $d_1$ & 0.01 & mm$^{-3}$ days$^{-1}$ & \cite{Kuznetsov1994, DOnofrio2005, DePillis2005, DOnofrio2012} \\
 Effectors death rate                                    & $d_0$ & 0.37 & days$^{-1}$ & \cite{Kuznetsov1994, DOnofrio2005, Su2009, DOnofrio2012} \\
 Innate immune growth rate                         & $\sigma$ &  $0.13 \times 10^5$ & cells days$^{-1}$ & \cite{Kuznetsov1994, DOnofrio2005, DePillis2005, DOnofrio2012} \\
\hline
\end{tabular}
\caption{
{\bf Model parameters.} The effects of model parameters $B$ and $r$, i.e. functional tumor-associated vasculature and effector cell recruitment rate respectively, are investigated by a model sensitivity analysis.}
\label{tab:params}
\end{table}


\section*{Dynamical analysis of the model}
\label{sec: dynamicalanalysis}

In this section, we analyze the model's behaviour with respect to two parameters, namely the effector cell recruitment rate $r$ and functional tumor-associated vasculature $B$.

\subsection*{Fixed point analysis}
\label{sec: fixedpoint}

\begin{figure}[H]
\begin{center}
\centerline{\includegraphics[width=0.77\textwidth]{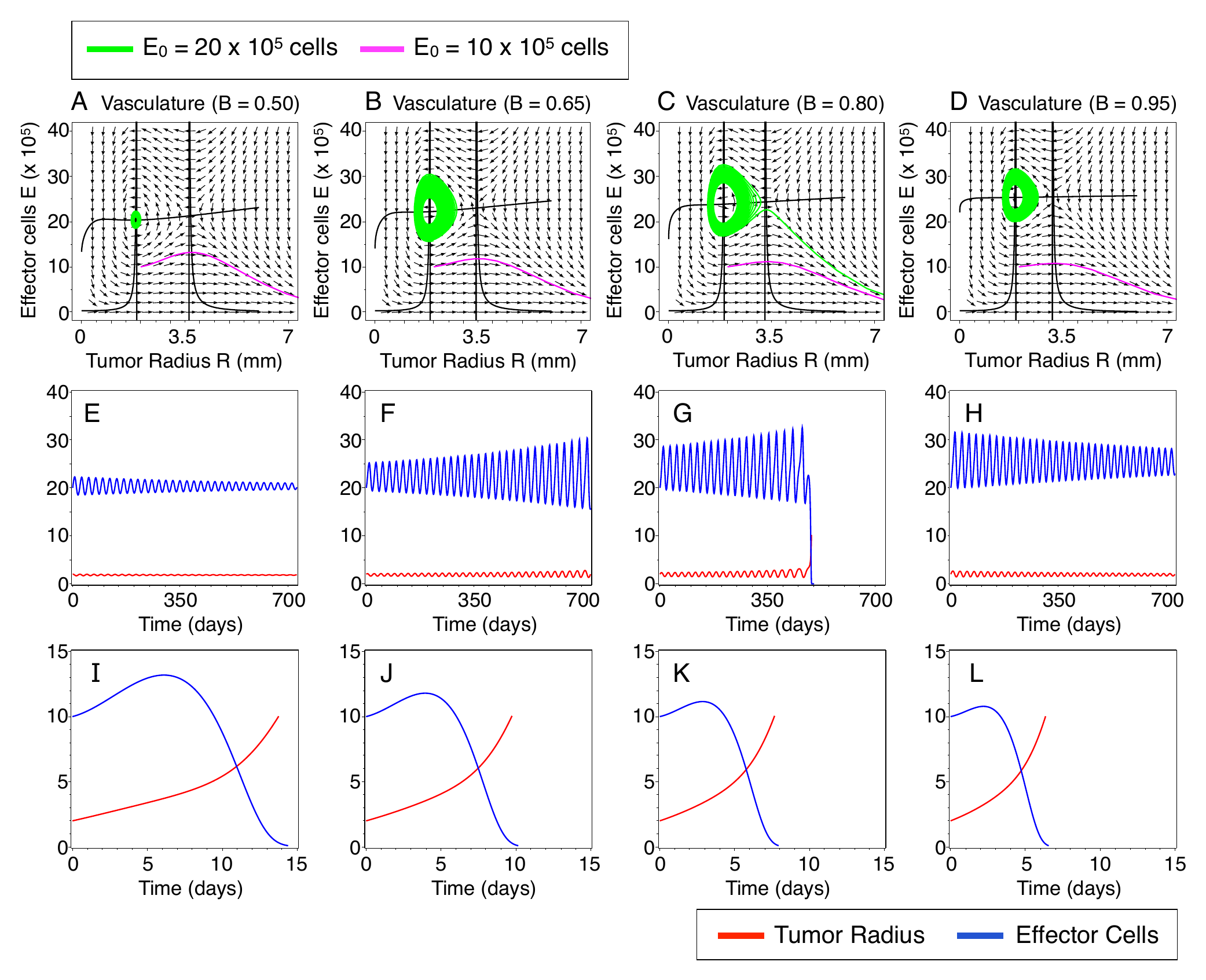}}
\caption{{\bf Phase portraits and long-term characteristic dynamics.} (A-D) Phase portraits of the tumor radius $R$ versus the concentration of effector cells $E$ for different functional levels of the tumor-associated vasculature $B$. The colored lines depict trajectories starting with the initial conditions $IC_1 = (R_0, E_0) = (2.0~\mbox{mm}, 20 \times 10^{5}~\mbox{cells})$ (green) and $IC_2 = (R_0, E_0) = (2.0~\mbox{mm}, 10 \times 10^{5}~\mbox{cells})$ (purple), respectively. The nullclines (zero-growth isoclines) of the dynamical system (black lines) are also plotted. (E-H) Temporal evolution of $R$ (red lines) and $E$ (blue lines) which correspond to the trajectories in panels (A-D) starting at the initial condition $IC_1$. (I-L) Temporal evolution of $R$ (red lines) and $E$ (blue lines) which correspond to the trajectories in panels (A-D) starting at the initial condition $IC_2$. The immune recruitment rate is $r = 0.57$~days$^{-1}$ and the remaining parameters are as in Tab.~\ref{tab:params}.}
\label{figure1}
\end{center}
\end{figure}

The first step towards analyzing the model dynamics is the identification of the fixed points along with their stability classification. Fig.~\ref{figure1}(A-D) depicts the phase portraits of the system of equations (\ref{eq:TE_1})-(\ref{eq:TE_2}) for different values of the model parameter $B$, while keeping $r$ constant and equal to $0.57$~days$^{-1}$. The black curves represent the nullclines, i.e. curves along which $dR/dt = 0$ and $dE/dt = 0$, and the colored curves the system trajectories for different initial conditions. The fixed points of the system are located at the intersection of the nullclines. In each case, we identify the existence of two fixed points corresponding to a low tumor radius ($\bold{L} = (R_L,E_L))$ and a high tumor radius ($\bold{H} = (R_H,E_H))$.

Fig.~\ref{figure1}(E-L) shows the evolution of the tumor radius and effector cells of the corresponding trajectories on the phase plain, presented in Fig.~\ref{figure1}(A-D). A quick glance at Fig.~\ref{figure1} reveals that the system trajectories initiated near the $\bold{L}$ fixed point follow an oscillatory behavior with a slowly varying amplitude that can be either increasing Fig.~\ref{figure1}(F,G) or decreasing Fig.~\ref{figure1}(E,H). This behavior indicates that, depending on the model parameter values, the $\bold{L}$ fixed point can be an attractor or a repellor. On the other hand, the trajectories initiated away from this point follow an exponential behavior which results in a boundless tumor growth while the amount of the effector cells fades out (see Fig.~\ref{figure1}(I-L)).

\begin{figure}[H]
\begin{center}
\centerline{\includegraphics[width=0.80\textwidth]{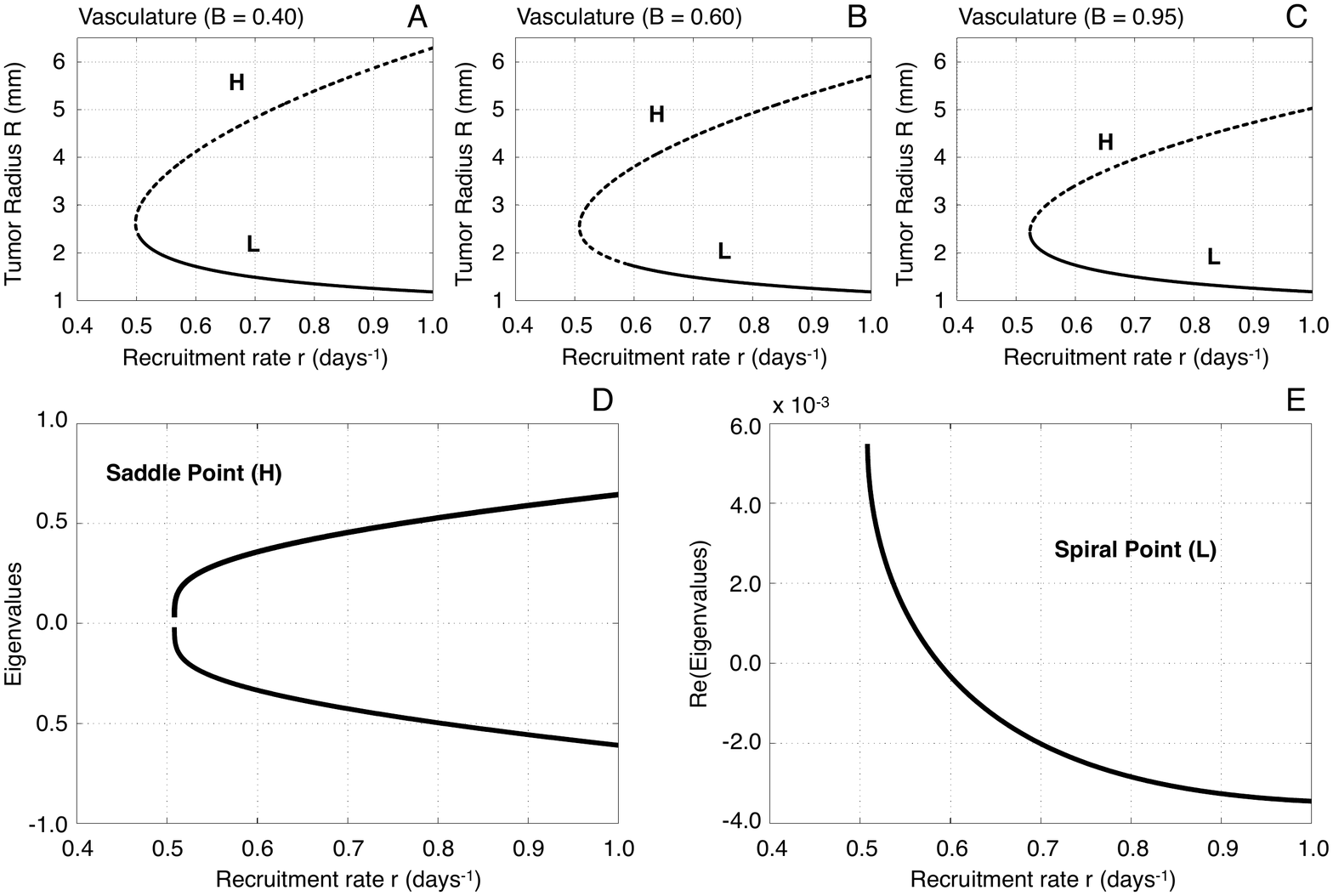}}
\caption{{\bf Bifurcation diagrams with respect to $r$ and local stability analysis.} (A-C) One-parameter bifurcation diagrams with respect to the effector cell recruitment rate $r$ for different values of the functional tumor-associated vasculature $B$ (0.40, 0.60 and 0.95, respectively). The upper branches correspond to the saddle point $\bold{H}$, whereas the lower branches to the spiral point $\bold{L}$. Solid lines depict stable fixed points and dotted lines the unstable fixed points. (D) Eigenvalues of the Jacobian estimated at the saddle point $\bold{H}$ with respect to the parameter $r$ for $B = 0.60$. (E) Real part of the eigenvalues of the Jacobian estimated at the spiral point $\bold{L}$ with respect to the parameter $r$ for $B = 0.60$.}
\label{figure2}
\end{center}
\end{figure}

To gain insight in the system behavior near the fixed points we construct the bifurcation diagrams with respect to the model parameter $r$ for different values of $B$ (see Fig.~\ref{figure2}(A-C)). The bifurcation diagrams were calculated by performing an arc-length continuation method. This method is a special case of numerical fixed-point continuation methods that ensures the continuation of solution branches at turning points \cite{kelley1999, keller1977numerical}.

For sufficiently small values of $r$, we obtain that there are no fixed points, which consequently implies that the tumor grows indefinitely. However, for a critical tumor radius $r_{cr}$, a homoclinic bifurcation occurs \cite{nayfeh2008,strogatz2001} giving rise to two states: a lower branch which corresponds to the $\bold{L}$ fixed point, and an upper branch which corresponds to the $\bold{H}$ fixed point. The local stability analysis shows that the $\bold{H}$ fixed point is a saddle point, and therefore unstable. On the other hand, the $\bold{L}$ fixed point is a spiral point which, depending on the value of the parameter $B$, can be unstable (spiral source) or stable (spiral sink) \cite{boyce1992elementary,iooss2012elementary}. Fig.~\ref{figure2}(D,E) shows the local stability analysis of $\bold{L}$ and $\bold{H}$ for $B = 0.60$, which corresponds to the bifurcation diagram in Fig.~\ref{figure2}(B).

\begin{figure}[H]
\begin{center}
\centerline{\includegraphics[height=0.35\textwidth]{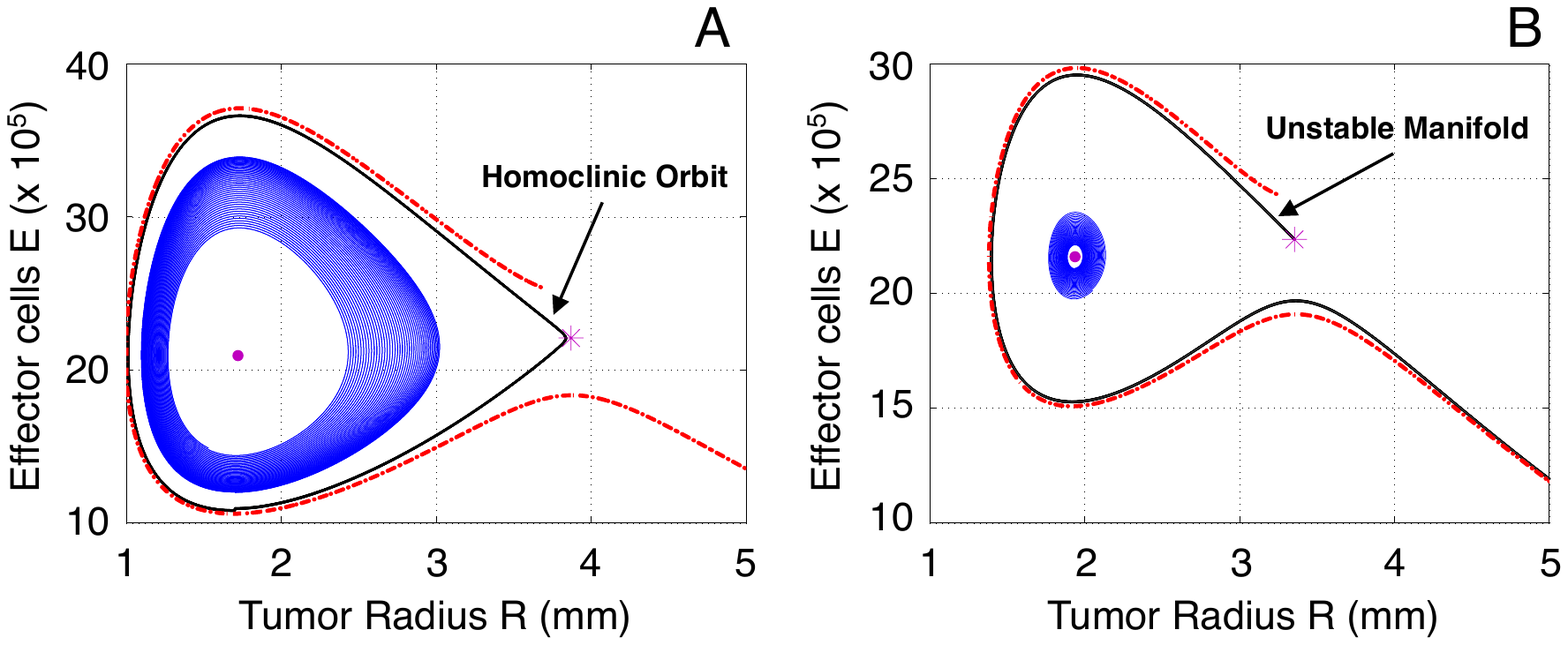}}
\caption{{\bf Classification of the system trajectories.} (A) Homoclinic orbit of the saddle point $\bold{H}$ (star) when the spiral point $\bold{L}$ (dot) is stable. (B) The unstable manifold of the saddle point $\bold{H}$ (star) when the spiral point $\bold{L}$ (dot) is unstable}
\label{figure3}
\end{center}
\end{figure}

In the case of $\bold{L}$ being a spiral sink, a system trajectory initiated inside the homoclinic orbit, i.e. the closed orbit which starts and returns to the saddle point in Fig.~\ref{figure3}(A), will follow regular oscillations with a slowly decreasing amplitude until it reaches the fixed point. This implies that the tumor radius will stay in a control bounded state. The homoclinic orbit defines the basin of attraction for the spiral sink \cite{nayfeh2008, strogatz2001}. Thus, any trajectory initiated outside the homoclinic orbit will result in an uncontrollable tumor growth while the concentration of effector cells fades out. On the other hand, if $\bold{L}$ is a spiral source, any initialization of the system close to $\bold{L}$ will produce regular oscillations with a slowly increasing amplitude. This behavior persists until the trajectory of the system reaches the unstable manifold of the saddle point $\bold{H}$ which drives the system towards an exponentially uncontrollable tumor evolution (see Fig.~\ref{figure3}(B)).

\begin{figure}[H]
\begin{center}
\centerline{\includegraphics[height=0.55\textwidth]{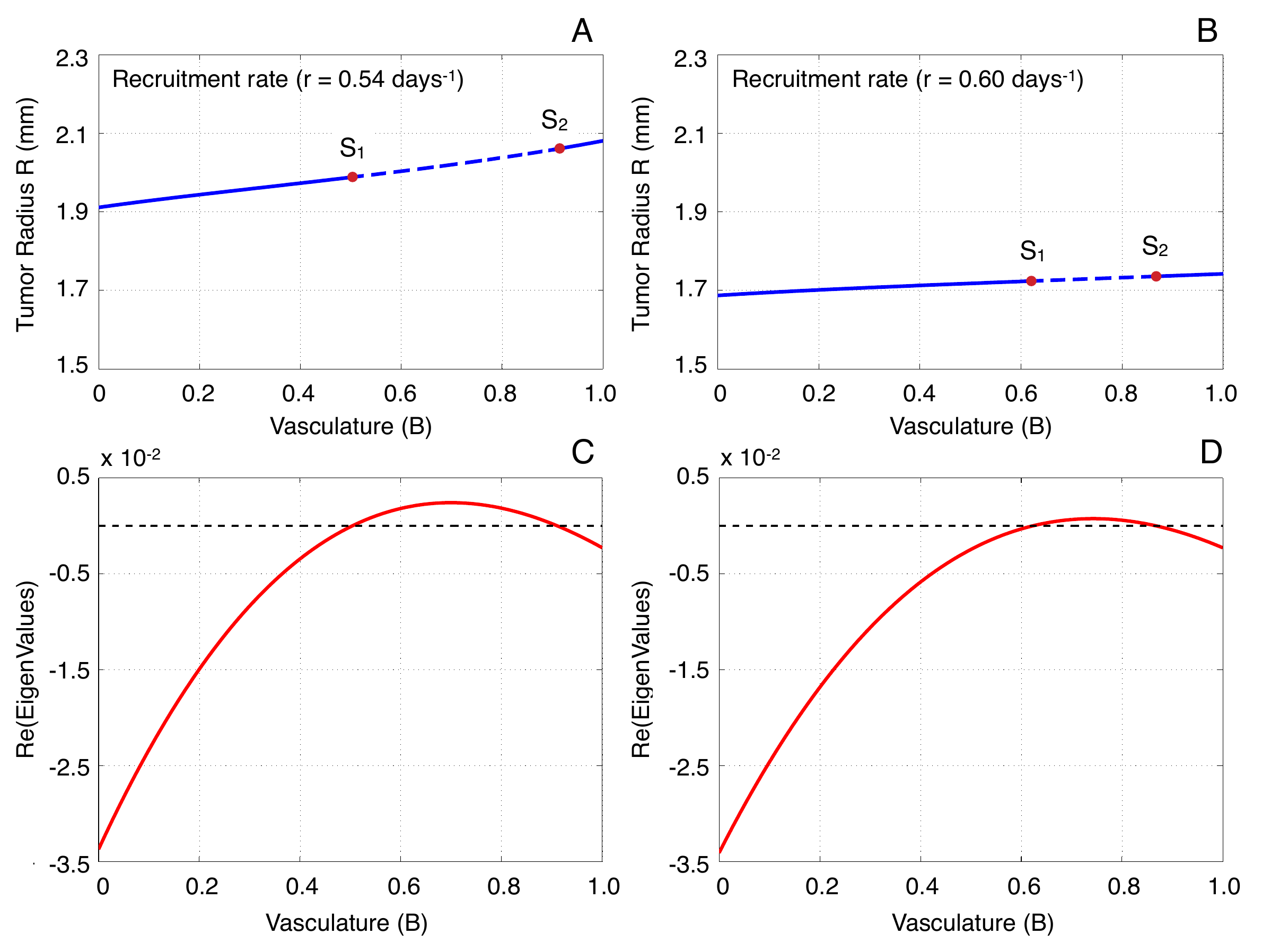}}
\caption{{\bf Stability analysis of the spiral point $\bold{L}$ with respect to parameter $B$.} (A,B) Stability analysis of $\bold{L}$ with respect to the functional tumor-associated vasculature $B$ for the immune recruitment rate $r$ equal to 0.54 and 0.60 days$^{-1}$, respectively. The labels $S_1$ and $S_2$ represent the bifurcation points, while the solid and dashed lines depict the stable and unstable solution branches, respectively. (C) Real part of the eigenvalues of the Jacobian estimated at the spiral point $\bold{L}$ with respect to $B$ for $r = 0.54$ days$^{-1}$. (D) Real part of the eigenvalues of the Jacobian estimated at the spiral point $\bold{L}$ with respect to $B$ for $r = 0.60$ days$^{-1}$. }
\label{figure4}
\end{center}
\end{figure}

\begin{figure}[H]
\begin{center}
\centerline{\includegraphics[height=0.30\textwidth]{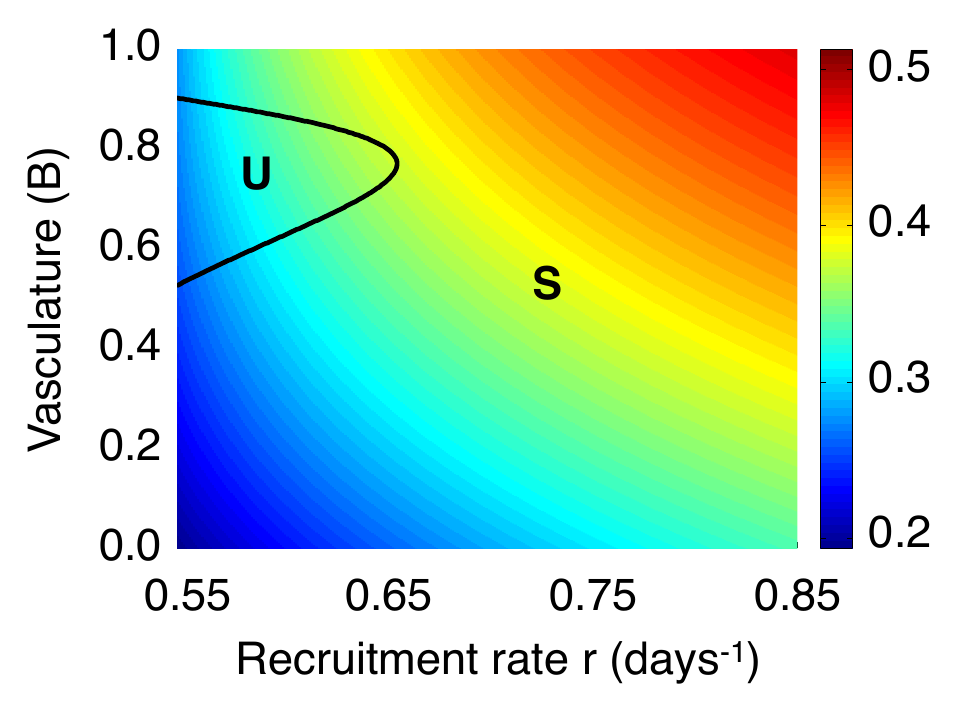}}
\caption{{\bf Frequency dependence on model parameters $B$ and $r$.} Imaginary part of the eigenvalues of the Jacobian estimated at the spiral point $\bold{L}$ with respect to the immune recruitment rate $r$ and the functional tumor-associated vasculature $B$. The labels ${\bf S}$ and ${\bf U}$ represent the regimes where the spiral point $\bold{L}$ is stable and unstable, respectively.}
\label{figure5}
\end{center}
\end{figure}

Interestingly, the time required for reaching the unstable manifold is significantly large. This system's behavior can be explained by performing a local stability analysis at the spiral point $\bold{L}$. Fig.~\ref{figure4}(A,B) illustrates the stability of $\bold{L}$ when $r$ is kept fixed and $B$ is allowed to vary. The eigenvalues of the Jacobian estimated at $\bold{L}$ are $\lambda=\mu \pm i \beta$. When $\mu < 0$, $\bold{L}$ is a spiral sink, while for $\mu > 0$, $\bold{L}$ is a spiral source. Fig.~\ref{figure4}(C,D) shows how $\mu$ changes with respect to parameter $B$. For all values of the parameters considered, we find that when $\bold{L}$ is a spiral source the real part of the eigenvalues is $\mu \ll 1$ with a maximum value of $\mu_{max}\cong 5.5 \times 10^{-3}$. Therefore, by setting $ \mu = \epsilon \ll 1$, we can identify two different time-scales describing the system's behavior: a fast time scale $t$ where the regular oscillations occur and a slow time scale $T=\epsilon t$ which describes the slowly varying amplitude. Fig.~\ref{figure5}(A) depicts the variation of $\beta$, which determines the oscillations frequency, with respect to the parameters $r$ and $B$. Notice that $\beta \in (0,1)$, i.e. $\beta \propto O(1)$.

A plausible question relates to the possibility of controlling tumor growth and with the conditions under which this control can be achieved. The local stability analysis reveals that, when $\bold{L}$ is a spiral sink, the homoclinic orbit creates a trapping region where the tumor remains controlled. However, when $\bold{L}$ is a spiral source, the tumor will finally escape innate immune control, albeit the fact that it will stay to a certain region for a long period of time. Therefore, a potential strategy aiming at controlling the tumor growth would be to keep the tumor near the $\bold{L}$ fixed point. In the next sections, we show how an external modulation of the immune system dynamics could be designed to limit the uncontrolled tumor growth in the case that $\bold{L}$ is a spiral source.


\section*{Mapping to a negatively damped harmonic oscillator via multiscale analysis}
\label{sec: multiscale}

The original model given by Eqs.~(\ref{eq:TE_1})-(\ref{eq:TE_2}) cannot be solved analytically and no control strategy can be easily designed. For this reason, we employ a multiscale approach to analyze the system's behavior near the spiral point $\bold{L}$. This approach, that efficiently describes the dynamics near a Hopf bifurcation point,  reveals the inherent multiscale structure of the problem by capturing the regular oscillations and constructing an envelope of the slowly varying amplitude in deterministic \cite{cole1996} or stochastic nonlinear systems \cite{klosek2005, kuske2003, kuske2007}. Thus, adopting such an approach we are able to map the initial system to a simplified negatively damped harmonic oscillator which allows to describe the self-sustained oscillations and the system's energy gain \cite{jenkins2013self}.

The system's behavior near the spiral source $\bold{L}$ can be described by linearizing the Eqs.~(\ref{eq:TE_1})-(\ref{eq:TE_2}) around $R = R_L$ and $E = E_L$. Then, we obtain a new system of equations approximating the evolution of the perturbations $u$ and $v$ given by:

\begin{equation}
u = R - R_L, \quad v = E - E_L,
\end{equation}

\noindent for $\left| u \right|\ll 1$ and $\left| v \right| \ll 1$. Then, the linearized system takes the following form:

\begin{equation}
\label{linearized}
\frac{d}{dt}\begin{pmatrix}
u  \\
v \end{pmatrix} \approx \bold{J}|_{L}  \begin{pmatrix} 
u\\
v \end{pmatrix} = \begin{bmatrix} \frac{\partial  F}{\partial  R}|_{L} & \frac{\partial  F}{\partial  E}|_{L} \\ &  \\ \frac{\partial  G}{\partial  R}|_{L} & \frac{\partial  G}{\partial  E}|_{L} \end{bmatrix}=\begin{bmatrix} a_1 & a_2\\ \\ a_3 & a_4 \end{bmatrix},
\end{equation}

\noindent where $\bold{J}|_{L}$ represents the Jacobian at the fixed point $\bold{L}$ and $F,G$ correspond to the right-hand side of Eqs.~(\ref{eq:TE_1})-(\ref{eq:TE_2}), respectively. The eigenvalues of the Jacobian are $\lambda= Tr(\bold{J}|_{L})/2 \pm \sqrt{(Tr(\bold{J}|_{L}))^2 - 4D}/2 = \mu \pm i \beta$. Moreover, $Tr(\bold{J}|_{L}) = a_1+a_4$ and $D = a_1 a_4-a_2 a_3$ are the trace and determinant of the Jacobian matrix, respectively. Since $\epsilon=\mu$ and $\beta \propto O(1)$, it follows that $a_1 \propto O(\epsilon)$, $a_4 \propto O(\epsilon)$, whereas $a_2 \propto O(1)$ and $a_3 \propto O(1)$. Moreover, we observe that $a_1 a_4 \propto O(\epsilon^2)$  which, consequently, results in $\beta =\sqrt{(a_1+a_4)^2-4(a_1 a_4-a_2 a_3)}/2 \approx \sqrt{-a_2 a_3}$. Thus, the eigenvalues can be approximated by $\lambda \approx \hat{\lambda} = \epsilon \pm i \omega$, where $\omega = \sqrt{- a_2 a_3}$. The condition $\epsilon \ll 1$ holds for each case where $\bold{L}$ is a spiral source and $a_2$ is always negative since it represents the death rate of tumor cells. Therefore, we can approximate the linearized system (\ref{linearized}) by the following one which incorporates the different order-terms:

\begin{equation}
\begin{aligned}
\label{modelapproximation}
\frac{d \hat{u}}{dt}=\epsilon\hat{u}+a_2 \hat{v}, \\
\frac{d \hat{v}}{dt}=a_3 \hat{u}+\epsilon \hat{v},
\end{aligned}
\end{equation}

\noindent where, $\hat{u}\approx u$ and $\hat{v}\approx v$. The approximate linearized system~(\ref{modelapproximation}) explicitly captures the different-order terms and allows to draw analogies with the field of Mechanics. More precisely, the system~(\ref{modelapproximation}) represents a perturbed harmonic oscillator. The unperturbed problem (i.e. when $\epsilon = 0$) is a linear harmonic oscillator with frequency $\omega$. The higher-order terms (or correction terms) insert small perturbations which result in a weakly negatively damped harmonic oscillator.

The solution of the approximate system~(\ref{modelapproximation}) can be expressed as:

\begin{equation}
\label{approximation}
\begin{pmatrix} 
\hat{u} \\
\hat{v} \end{pmatrix} =  A(T)  \begin{pmatrix} \frac{a_2}{\omega} \sin(\omega t) \\ \cos(\omega t) \end{pmatrix} + B(T)  \begin{pmatrix} \frac{a_2}{\omega} \cos(\omega t) \\ -\sin(\omega t) \end{pmatrix},
\end{equation}

\noindent where $A(T)$ and $B(T)$ contain the information regarding the slowly varying amplitude depending on $T$, where $T=\epsilon t$ for $\epsilon \ll 1$. The multiscale assumption is that the functions $A(T)$ and $B(T)$ evolve at the slow time scale $T$ and are considered to be constant with respect to the oscillations with frequency $\omega$, evolving on the fast time scale $t$. Notice that the variables $T=\epsilon t \propto O(\epsilon)$ and $t \propto O(1)$ are considered to be independent. The constant term $a_2/\omega$ in the expression of $\hat{u}$ has been used to simplify the upcoming calculations.

In order to estimate the functions $A(T)$ and $B(T)$, we assume that they follow evolution equations of the form:

\begin{equation}
\label{amplitude}
dA=f_1 dT, \quad dB=f_2 dT.
\end{equation}

\noindent Then, by taking the differentials of the system of equations in~(\ref{approximation}), we have that:

\begin{equation}
\begin{aligned}
\label{differentials}
& d \hat{u}=\frac{\partial \hat{u}}{\partial t} dt+ \frac{\partial \hat{u}}{\partial A} dA+\frac{\partial \hat{u}}{\partial B}dB \\
& \hspace{4mm} = a_2\hat{v} \hspace{0.5mm} dt+ \left( f_1\frac{a_2}{\omega} \sin(\omega t)+f_2\frac{a_2}{\omega} \cos(\omega t)\right)dT, \\
& d \hat{v}=\frac{\partial \hat{v}}{\partial t} dt+ \frac{\partial \hat{v}}{\partial A}dA+\frac{\partial \hat{v}}{\partial B}dB \\
& \hspace{4mm} =a_3\hat{u} \hspace{0.5mm} dt+ \left( f_1 \cos(\omega t)-f_2 \sin(\omega t)\right)dT.
\end{aligned}
\end{equation}

\noindent Notice that the system of equations in~(\ref{modelapproximation}) and~(\ref{differentials}) are equivalent. Therefore, by equating the terms which are of the same order, we obtain that:

\begin{align}
\label{equate1} f_1\frac{a_2}{\omega} \sin(\omega t)+f_2\frac{a_2}{\omega} \cos(\omega t) &= A(T)\frac{a_2}{\omega} \sin(\omega t)+B(T)\frac{a_2}{\omega} \cos(\omega t), \\
\label{equate2} f_1 \cos(\omega t)-f_2 \sin(\omega t) &= A(T)\cos(\omega t)-B(T) \sin(\omega t).
\end{align}

\noindent To estimate the functions $f_1$ and $f_2$, we project the Eqs.~(\ref{equate1})-(\ref{equate2}) onto the fast dynamics. This is an important step of the calculations in order to isolate the amplitude of functions $f_1$ and $f_2$ \cite{kuske2007,kuske2003}. For instance, multiplying Eq.~(\ref{equate1}) by $\sin(\omega t)$ and integrating over a period of time $[0,2\pi/\omega]$ , we have that:

\begin{align}
& \int_{0}^{2\pi/\omega} \left( f_1\frac{a_2}{\omega} \sin(\omega t)+f_2\frac{a_2}{\omega} \cos(\omega t)\right) \sin(\omega t) \hspace{0.5mm} dt \nonumber \\
& =\int_{0}^{2\pi/\omega} \left(A(T)\frac{a_2}{\omega} \sin(\omega t)+B(T)\frac{a_2}{\omega} \cos(\omega t)\right) \sin(\omega t) \hspace{0.5mm} dt, \nonumber
\end{align}

\noindent which results in $f_1 = A$. In a similar way, multiplying Eq.~(\ref{equate1}) by $\cos(\omega t)$ and following the same procedure, we find $f_2 = B$. Consequently, the solution of the approximate linearized system~(\ref{modelapproximation}) is given by:

\begin{equation}
\label{approximationanalytic}
\begin{pmatrix} 
\hat{u} \\
\hat{v} \end{pmatrix} =  A_0 e^{T}  \begin{pmatrix} \frac{a_2}{\omega} \sin(\omega t) \\ \cos(\omega t) \end{pmatrix} + B_0 e^{T}  \begin{pmatrix} \frac{a_2}{\omega} \cos(\omega t) \\ -\sin(\omega t) \end{pmatrix}, 
\end{equation}

\noindent where the positive parameters $A_0$ and $B_0$ are defined by the initial conditions of the original model (\ref{eq:TE_1})-(\ref{eq:TE_2}). In addition, Eq.~(\ref{approximationanalytic}) can be further simplified to take the following form:

\begin{equation}
\label{approximationsimplified}
\begin{pmatrix} 
\hat{u} \\
\hat{v} \end{pmatrix} =  \begin{pmatrix} M e^T \sin(\omega t+\phi) \\ N e^T \sin(\omega t+\delta) \end{pmatrix},
\end{equation}

\noindent where $M \cos(\phi)=A_0 \hspace{0.5mm} a_2/\omega$, $M \sin(\phi)=B_0 \hspace{0.5mm} a_2/\omega$, $N \sin(\delta)=A_0$ and $N \cos(\delta)= -B_0$. We show that the functions $\hat{u}$ and $\hat{v}$ are orthogonal since:

\begin{align}
\label{orthogonal}
\cos(\delta-\phi)=\cos(\delta)\cos(\phi)+\sin(\delta)\sin(\phi) \\
=\frac{-B_0}{N} \frac{A_0 }{M} \frac{a_2}{\omega}+ \frac{A_0}{N} \frac{B_0}{M}\frac{a_2}{\omega}=0 \\
\Leftrightarrow \delta=\phi+\frac{\pi}{2}.
\end{align}
 
\noindent Then, the system of the approximate solutions becomes:
 
 \begin{equation}
\label{approximationsimplifiednew}
\begin{pmatrix} 
\hat{u} \\
\hat{v} \end{pmatrix} =  \begin{pmatrix} M e^T \sin(\omega t+\phi) \\ N e^T \cos(\omega t+\phi) \end{pmatrix}.
\end{equation}

The advantage of using the proposed approach is that the solutions of the approximate model~(\ref{approximationsimplifiednew}) depend directly on experimentally accessible parameters and the initial conditions. We focus now on the stability properties of the system~(\ref{modelapproximation}) with respect to the time evolution of the variables $\hat{u}$ and $\hat{v}$. To that end, we construct a Lyapounov functional as:

\begin{equation}
\label{energy}
V\left(\hat{u},\hat{v}\right)=\frac{1}{2}\left(a_3 \hat{u}^2-a_2 \hat{v}^2\right),
\end{equation}

\noindent where $V$ is always positive definite since $a_2 < 0$, and the time-derivative of $V$ is given by:

\begin{equation}
\label{energyrate}
\begin{aligned}
&\frac{dV}{dt}=a_3 \hat{u} \frac{d\hat{u}}{dt}-a_2 \hat{v} \frac{d\hat{v}}{dt} \\
&\hspace{2.7mm} = \epsilon \left(a_3 \hat{u}^2-a_2 \hat{v}^2 \right)+\left(a_3 a_2 \hat{u} \hat{v}-a_3 a_2 \hat{u} \hat{v}\right)\\
&\hspace{2.7mm} = \epsilon \left(a_3 \hat{u}^2-a_2 \hat{v}^2 \right) > 0.
\end{aligned}
\end{equation}

\noindent Since $dV/dt > 0$ the system gains energy. The average energy $\langle V \rangle$ over a period is estimated by using the form of the approximate solutions~(\ref{approximationsimplified}) and considering the multiscale assumption:

\begin{equation}
\label{averageenergy}
\langle V \rangle = \int_{0}^{2\pi/\omega} V \hspace{0.5mm} dt = \frac{1}{2}\left(a_3 \hat{u}^2-a_2 \hat{v}^2\right)dt \approx e^{2 T} \frac{1}{4}\left(a_3 M^2- a_2 N^2 \right).
\end{equation}

\noindent The term $\frac{1}{4}\left(a_3 M^2- a_2 N^2 \right)$ is constant and depends on time-invariant parameter values and the initial conditions. Consequently, the average energy gain rate over a period is equal to:

\begin{equation}
\label{averarateenergy}
\frac{d\langle V \rangle}{dt}=2 \epsilon \langle V \rangle.
\end{equation}

The previous results suggest that a therapeutic strategy, which influences the effector cell dynamics, should be designed in a way to {``}pump out energy" with an average rate which is greater than the system's gain energy. In particular, the per period gain rate is equal to $2 \epsilon$ according to relation~(\ref{averageenergy}). Thus, if we introduce an external immune-modulatory term $h(\hat{v})$ to intervene in the tumor-effector cell interactions, the system of equations~(\ref{modelapproximation}) becomes:

\begin{equation}
\begin{aligned}
\label{interfer}
& \frac{d \hat{u}}{dt}=\epsilon \hat{u}+a_2 \hat{v}, \\
& \frac{d \hat{v}}{dt}=a_3 \hat{u}+\epsilon \hat{v} + h(\hat{v}).
\end{aligned}
\end{equation}

\noindent According to Eq.~(\ref{energyrate}), the rate $dV/dt$ is at most of the order of $O(\epsilon)$. Therefore, the function $h(\hat{v})$ should be of the order of $O(\epsilon)$, for instance:

\begin{equation}
\label{control}
h(\hat{v})=\epsilon \hat{z}(t).
\end{equation}

In order to calculate the function $\hat{z}(t)$, we require that the energy $V_h$ of the system~(\ref{interfer}) should have a negative rate, that is: 
 
\begin{equation}
\label{newenergyrate}
 \begin{aligned}
&\frac{dV_{h}}{dt} =a_3 \hat{u} \frac{d\hat{u}}{dt}-a_2 \hat{v} \frac{d\hat{v}}{dt} \\
& = \epsilon \left(a_3 \hat{u}^2-a_2 \hat{v}^2 -a_2 \hat{v} \hat{z}(t) \right)+\left(a_3 a_2 \hat{u} \hat{v}-a_3 a_2 \hat{u} \hat{v}\right)\\
& = \epsilon \left(a_3 \hat{u}^2-a_2 \hat{v}^2-a_2 \hat{v} \hat{z}(t) \right) \leq 0.
\end{aligned}
\end{equation}

\noindent The negative energy rate means that a trajectory will flow {``}downhill" towards a stable fixed point. In the limit $dV_{h}/dt = 0$, we find an expression for the \textit{zero-rate energy function} $\hat{z}(t)$ in terms of the solutions $\hat{u}$ and $\hat{v}$, given by:

\begin{equation}	
\label{controlsolution}
\hat{z}(t) = \frac{a_3 \hat{u}^2-a_2 \hat{v}^2}{a_2 \hat{v}}.
\end{equation}

\noindent Substituting the solutions of $\hat{u}$ and $\hat{v}$ from Eq.~(\ref{approximationsimplifiednew}), we have an analytical expression of the zero-rate energy function $\hat{z}(t)$, that is:

\begin{equation}
\label{controlsolutionexpressed}
\hat{z}(t) = \frac{a_3}{a_2}\frac{M^2}{N} e^{T} \tan(\omega t + \phi)\sin(\omega t + \phi)-M e^{T} \cos(\omega t + \phi).
\end{equation}

\noindent The function $\hat{z}(t)$ has singularities at the points where the function $\cos(\omega t + \phi)$ is equal to zero, i.e. $t_{sing}=\left(\kappa \pi +\frac{\pi}{2}-\phi \right)/ \omega$, $\kappa=0,1,2,...$. However, the effect of $\hat{z}(t)$, i.e. the integral of the function within a small time interval $\left[ t_1,t_2 \right]\subset\left[0,2\pi/\omega \right]$, is finite.

\begin{figure}[H]
\begin{center}
\centerline{\includegraphics[height=0.30\textwidth]{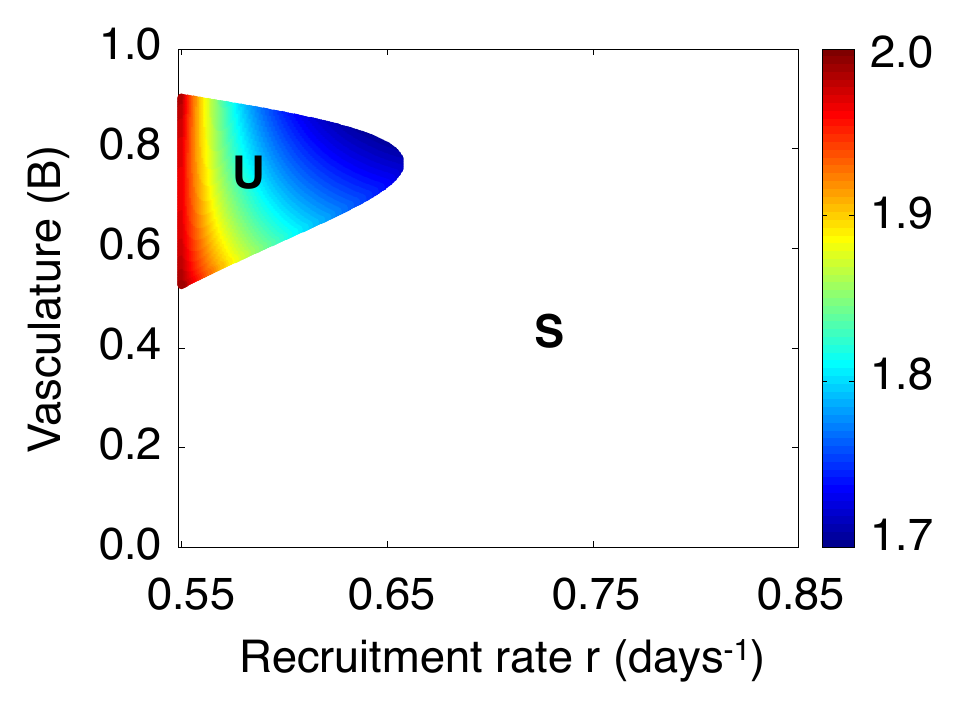}}
\caption{{\bf Dependence of the zero-rate energy function $\hat{z}(t)$ effect on model parameters $B$ and $r$.} Dependence of the effect of $\hat{z}(t)$ on the immune recruitment rate $r$ and functional tumor-associated vasculature $B$ in the unstable regime of $\bold{L}$ denoted by $\bold{U}$. The label $\bold{S}$ stands for the regime where the spiral point is stable and tumor control is reached.}
\label{figure6}
\end{center}
\end{figure}

Fig.~\ref{figure6} illustrates the effect of $\hat{z}(t)$ by integrating between $t = 0$ and $t = \pi/\omega$, for values of the model parameters $r$ and $B$ where $\bold{L}$ is unstable. In each case, the system was initialized near the spiral point $\bold{L}$. 

Using the function $\hat{z}(t)$ is not feasible for practical purposes. However, we can design meaningful therapy functions $\hat{z}_{ext}(t)$ that induce the same or greater effects in a certain time interval [$t_1$, $t_2$] as the function $\hat{z}(t)$, that is:
 
 \begin{equation}
\label{integraleffectexternal}
\int_{t_1}^{t_2} \hat{z}(t) \hspace{0.5mm} dt  \leq  \int_{t_1}^{t_2} \hat{z}_{ext}(t) \hspace{0.5mm} dt.
\end{equation}
 
This will result in {``}pumping out energy" at a rate greater than the system's energy gain. In the next section, we estimate the function $\hat{z}(t)$ which represents the zero-rate energy scenario, as well as we present numerical simulation results for specific values of the model parameters. Furthermore, based on the behavior of the function $\hat{z}(t)$, we suggest an efficient external immune-modulatory term $\hat{z}_{ext}(t)$ to fulfill the relation~(\ref{integraleffectexternal}) which results in long-term tumor control.


\section*{Results: theory-driven therapeutic design}
\label{sec: simulation}

In this section, we design an immuno-therapeutic proposal derived from our model analysis. We first compare the approximate solutions with those obtained from the original model~(\ref{eq:TE_1})-(\ref{eq:TE_2}). Then, we show how the defined energy function in Eq.~(\ref{energy}) for the approximate linearized system can be used to describe the system's energy gain. Finally, we design an external immuno-modulatory strategy following the behavior of the zero-rate energy function given in Eq.~(\ref{controlsolutionexpressed}). 

To illustrate the simulation results and without loss of generality, we select the following parameter values: $r = 0.6$ days$^{-1}$ and $B = 0.8$, which result in $\epsilon=5.85 \cdot 10^{-4}$ and $\omega=\sqrt{-a_2 a_3} = 0.336$ rad/days. Notice that similar results can be obtained for each parameter set where $\bold{L}$ is a spiral source, since $\epsilon \ll 1$ always holds. For numerical integration, we consider a $4^{th}$ order Runge-Kutta method. The time step was set equal to $dt = 0.001$ days which is sufficiently small.

\subsection*{Validity of approximate solution}

\begin{figure}[H]
\begin{center}
\centerline{\includegraphics[width=0.75\textwidth]{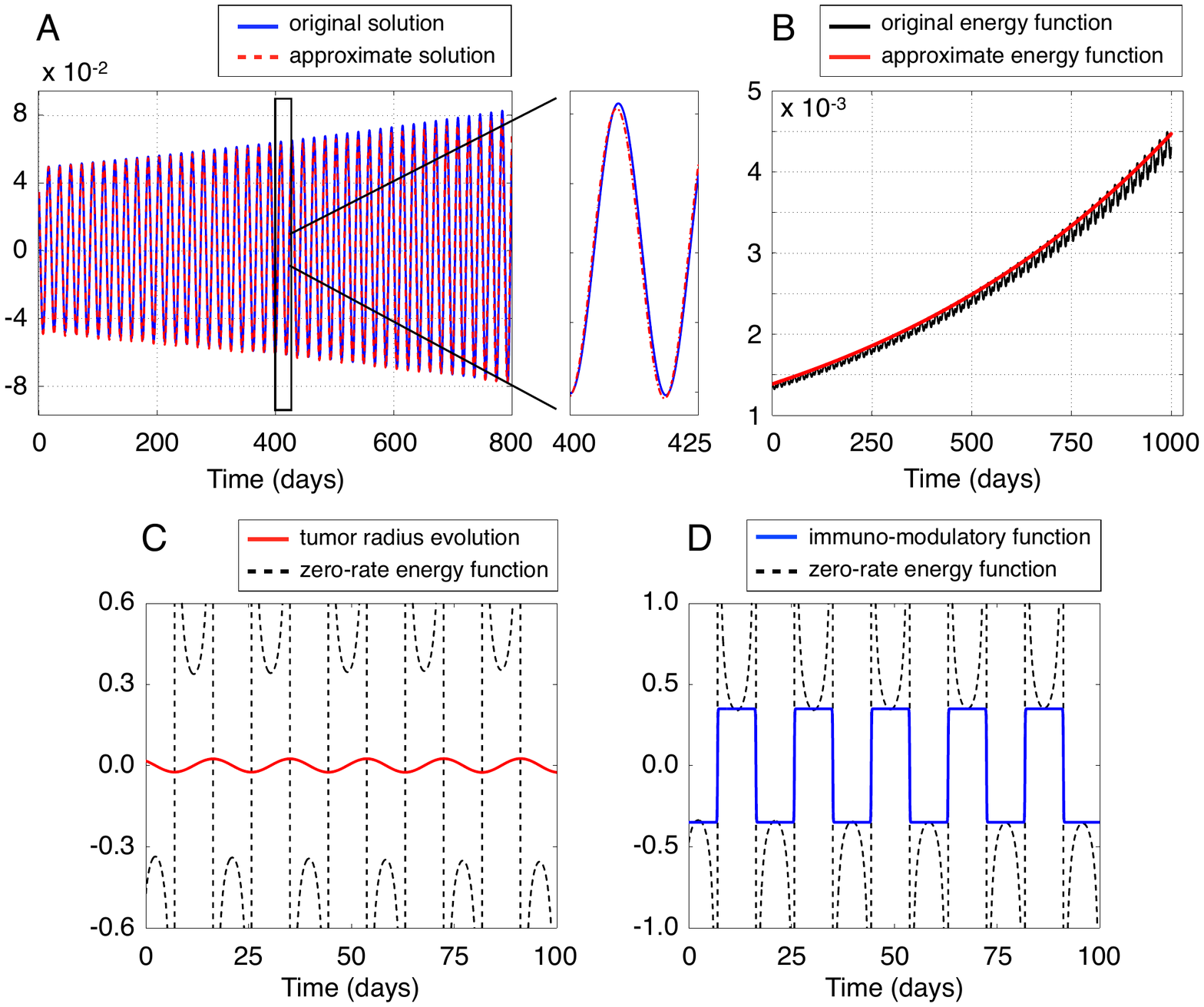}}
\caption{{\bf Validation of approximate solutions.} (A) Evolution of the approximate perturbation $\hat{u}$ compared to the original model perturbation $u = R - R_L$. (B) Evolution of the energy function of the approximate linearized model defined in Eq.~(\ref{energy}) against the energy estimated by using the original model~(\ref{eq:TE_1})-(\ref{eq:TE_2}). (C) The zero-rate energy function in Eq.~(\ref{controlsolution}) estimated by using the approximate solution of the system~(\ref{orthogonal}). (D) The immuno-modulatory function $\hat{z}_{ext}(t)$ for $\alpha = - 0.6$ and $\gamma=100$ compared with the zero-rate energy function $\hat{z}(t)$.}
\label{figure7}
\end{center}
\end{figure}

At this point, we need to validate the performance of the approximate solutions. Fig.~\ref{figure7}(A) depicts the evolution of the approximate perturbation $\hat{u}$ compared to the perturbation $u = R - R_L$ of the initial model given by Eqs.~(\ref{eq:TE_1})-(\ref{eq:TE_2}), referred in what follows as original model perturbation. The approximate solutions $\hat{u}$ were estimated by performing a numerical integration of the system~(\ref{modelapproximation}), while $u = R - R_L$ by numerically integrating the original model~(\ref{eq:TE_1})-(\ref{eq:TE_2}). Both systems of equations were initialized close to the spiral point $\bold{L}$ and the simulation time was set $t_{sim} = 1000$ days. A quick glance at Fig.~\ref{figure7}(A) reveals that the solution derived by the approximate model~(\ref{modelapproximation}) is very close to the original one, which demonstrates the validity of the approach. Moreover, Fig.~\ref{figure7}(A) shows a comparison of the original and approximate model solutions by zooming in the time interval between days 400 and 425. In this interval, the maximum error between such solutions was found to be of the order of $10^{-3}$.

Fig.~\ref{figure7}(B) compares the evolution of the approximate energy function defined in Eq.~(\ref{energy}) against the energy of the original model (\ref{eq:TE_1},\ref{eq:TE_2}). As in the previous case, the energy function of the approximation is in a good agreement with that obtained from the original model. Therefore, we expect that an external immune-modulatory strategy, which results in a negative energy rate, can be used with the same efficacy to the original model.

\subsection*{Construction of the therapy term}

Fig.~\ref{figure7}(C) depicts the zero-rate energy function in Eq.~(\ref{controlsolution}) estimated by using the approximate solutions $\hat{u}$ and $\hat{v}$ from the system~(\ref{interfer}). As we observe, this function diverges at the singularity points, which means that it cannot be used directly to induce tumor control.

It should be noted that, the zero-rate energy function changes sign according to the evolution of the variable $u=R-R_L$, i.e. the function follows the monotonicity of tumor evolution. Hence, an external immuno-modulatory therapy $\hat{z}_{ext}(t)$ that satisfies the condition~(\ref{integraleffectexternal}) should follow the evolution of the tumor by increasing the recruitment rate of effector cells when the tumor radius increases and decreasing the immune recruitment rate when the tumor radius decreases. A simple approximation of the therapy function would be a step function having a constant value and the same sign to the zero-rate energy function at the same time interval:

\begin{equation}
\label{external}
\hat{z}_{ext}(t) = \alpha \tanh(\gamma \hat{v}),
\end{equation} 
 
\noindent where $\gamma \gg 1$ expresses the sigmoidal steepness, and the selection of $\alpha$ should fulfill the relation (\ref{integraleffectexternal}). Fig.~\ref{figure7}(D) depicts the immuno-modulatory function $\hat{z}_{ext}(t)$ for $\alpha = -0.6$ and $\gamma=100$, compared with the zero-rate energy function $\hat{z}(t)$ estimated.

\subsection*{Performance of the suggested therapy}

The application of the external therapy in Eq.~(\ref{external}) implies the addition of a new term in Eqs.~(\ref{eq:TE_1})-(\ref{eq:TE_2}), that is:

\begin{eqnarray}
\label{eq:TE_new1} \frac{dR}{dt} = \frac{1}{3} (\lambda_M B - \lambda_A) R + \lambda_M (1 - B) L_D \left( \frac{1}{\tanh(R / L_{D})} - \frac{L_{D}}{R} \right) - c E R f(R,B), \\
\label{eq:TE_new2} \frac{dE}{dt} =  r \frac{R^3}{K + R^3} E - d_1 E R^{3} f(R,\alpha) - d_0 E + \sigma + \epsilon \hspace{0.5mm} \alpha \tanh(\gamma v(t)),
\end{eqnarray}

\noindent where $v(t)=E(t) - E_L$.

\begin{figure}[H]
\begin{center}
\centerline{\includegraphics[width=0.85\textwidth]{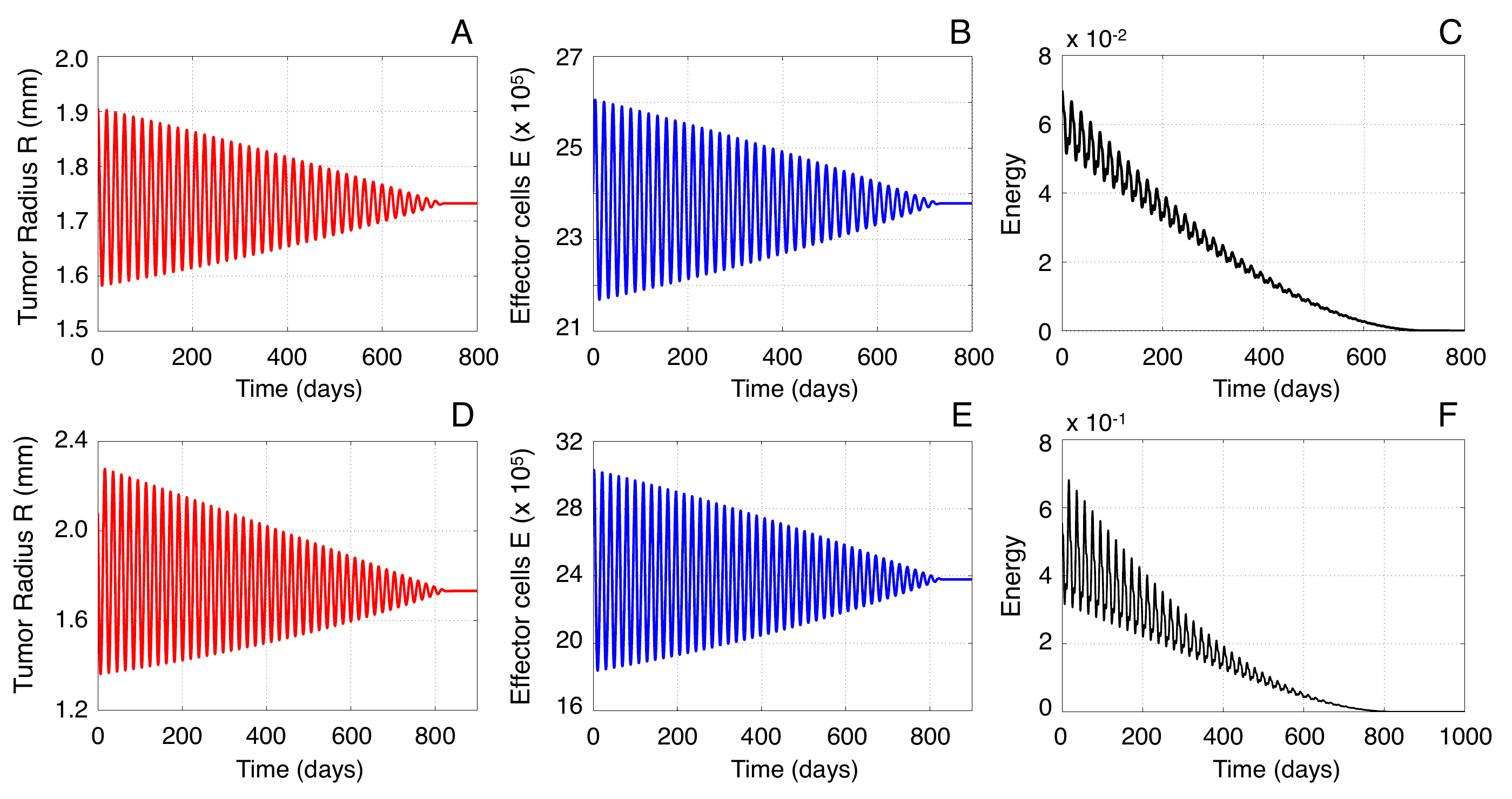}}
\caption{{\bf Performance of the proposed therapy term.} Evolution of the tumor radius (A) and concentration of effector cells (B) by considering the external immuno-modulatory function~(\ref{external}) with parameters $\alpha= -0.6$ and $\gamma=100$. (C) Energy of the system of Eqs.~(\ref{eq:TE_new1})-(\ref{eq:TE_new2}) with $\alpha= -0.6$ and $\gamma=100$. Evolution of the tumor radius (D) and concentration of effector cells (E) by initializing the system away from the spiral point $\bold{L}$ and considering the external immuno-modulatory function with $\alpha= -25$ and $\gamma=100$. (F) Energy of the system of Eqs.~(\ref{eq:TE_new1})-(\ref{eq:TE_new2}), with $\alpha= -25$ and $\gamma=100$, initialized away from the spiral point $\bold{L}$.}
\label{figure8}
\end{center}
\end{figure}

Fig.~\ref{figure8}(A,B) respectively shows the evolution of the tumor radius and effector cells by numerically integrating Eqs.~(\ref{eq:TE_new1})-(\ref{eq:TE_new2}) which include the proposed external immuno-modulatory function~(\ref{external}). The system is initialized near the spiral point $\bold{L}$ and parameter values of the external function were selected equal to $\alpha=-0.6$ and $\gamma=100$. In doing so, we obtain that the system reaches a stable steady state, i.e. the tumor remains controlled. Fig.~\ref{figure8}(C) illustrates how the energy of the system of Eqs.~(\ref{eq:TE_new1})-(\ref{eq:TE_new2}) evolves. The energy decreases with time and becomes zero as the system reaches the steady state. It is worth pointing out that $\alpha$ was selected not only to satisfy the relation~(\ref{integraleffectexternal}), but also to result in the system's fast convergence to a steady state.

Fig.~\ref{figure8}(D,E) respectively shows the evolution of the tumor radius and concentration of effector cells by numerically integrating Eqs.~(\ref{eq:TE_new1})-(\ref{eq:TE_new2}) when the system is initialized away from the spiral point $\bold{L}$. In this case, the approximate solution is expected to deviate from the solutions of the original model. Fig.~\ref{figure8}(F) represents the temporal evolution of the corresponding energy. The parameter $\alpha$ was selected to be equal to $-25$ to fulfil the relation~(\ref{integraleffectexternal}), as well as to provide a fast convergence to a steady state. Interestingly enough, in this case the system also reaches a steady state, even though the approximate system is not accurate enough. Consequently, the proposed external immuno-modulatory function is shown to be adequate in controlling tumor growth, even when the system is initialized far to the spiral point $\bold{L}$.


\section*{Discussion}

In this article, we investigate the therapeutic potential of immunomodulatory interventions against tumor growth. To that end, we consider a model introduced in \cite{Hatzikirou2015} that describes the dynamic interplay between vascularized tumor growth and effector cell responses. Our goal is to identify an external modification of effector cell dynamics that allows for controlling tumor growth. With the help of bifurcation analysis, we identified a unstable oscillatory regime that induces tumor evasion from immuno-surveillance. The characteristic feature of this regime is the oscillations occur at a faster time scale than the amplitude dynamics. Exploiting this time scale separation and via temporal multiscale analysis, we map our model onto a weakly negatively damped harmonic oscillator. This approximation allowed us to identify an analytical expression for the additive control term to the effector cell dynamics. This term acts as an external immunomodulatory therapy where the numerical simulations evidence its efficacy in controlling tumor growth. 

The crucial question concerns the translational potential of our theory-driven therapeutic proposal into clinical practice. Our suggested immunomodulatory strategy is relevant to small enough, non-invasive tumors that are initially controlled by the immune system but eventually  evade immuno-surveillance. The latter occurs when tumors exceed a critical size where immune responses are unable to confer any control. As stated above, our model suggests that such tumor evasion may take place in the form of oscillations of slowly increasing amplitude. The proposed therapeutic strategy is based on the synchronization of immuno-stimulating  and -suppressive phases with tumor growth dynamics. Although immuno-stimulating therapies seem to be expected and plausible \cite{Ghirelli2013, Rosenberg2014}, immuno-suppression sounds counter-intuitive and dangerous. However, the latter occurs in clinical practice during chemotherapeutic interventions \cite{Zitvogel2008}. Therefore, a potential realization of our proposed strategy could be mediated by a combination of state-of-the-art immunotherapies \cite{Zou2005, Finn2012, Fridman2012, Rosenberg2014} and chemotherapeutic modules. The latter not only will play the role of immuno-suppressor, but also will slow down tumor growth dynamics. Needless to state that such a therapeutic suggestion requires experimental validation and further investigation.

Finally, we conclude by pointing out the limitations of the present work. Paramount among them, introducing vascularization dynamics should be not only a natural, but also an insightful extension of the proposed model. At this state tumor vascularization $B$ is considered as a constant parameter in time. Thus, making vascularization dynamic would make sense to model further hypoxic effects, such as necrosis or hypoxic-induced invasion. Moreover, the immune system is much more complicated than its description in the current model, involving much more cell types and interactions. In particular, immune system is often regarded as acting to suppress tumor growth, however it is now clear that it can be both stimulatory and inhibitory, as in the case of tumor-associated macrophages \cite{Visser2006, Mantovani2008}. Moreover, our model needs be extended for invasive tumors. Nevertheless, mathematical modeling can help to improve our understanding of the interplay between the several competing factors that have complex implications for cancer therapy. We hope that the model presented here could provide a step towards this direction.


\section*{Acknowledgments}
A. I. Reppas, J. C. L. Alfonso and H. Hatzikirou gratefully acknowledge the support of the German Excellence Initiative via the Cluster of Excellence EXC 1056 Center for Advancing Electronics Dresden (cfAED) at the Technische Universit\"{a}t Dresden. J. C. L. Alfonso and H. Hatzikirou also acknowledge the German Federal Ministry of Education and Research (BMBF) for the eMED project SYSIMIT (01ZX1308D).


\bibliographystyle{ieeetr}

\end{document}